\documentclass{article}
\usepackage[english]{babel}
\usepackage{amsmath}
\usepackage{amsfonts}
\usepackage{amssymb}
\usepackage{graphicx}

\providecommand{\U}[1]{\protect\rule{.1in}{.1in}}
\graphicspath{    {converted_graphics/}    {/}}

\begin{document}

\title{Evaluating the Force Concept Inventory for different student groups
at the Norwegian University of Science and Technology}
\author{J.R. Persson \\
Programme for Teacher Education\\
Norwegian University of Science and Technology\\
NO-7491 Trondheim\\
Norway\\
}
\maketitle

\begin{abstract}
The Force Concept Inventory (FCI) was developed by Hestenes, Wells and
Swackhamer, in order to assess student understanding of the concept of
force. FCI has been used for over 20 years and in different countries. When
applying the inventory in a new context it is important to evaluate the
reliability and discrimination power of this assessment tool. In this study
the reliability and discrimination power are evaluated in the context of
Engineering education at a Norwegian university, using statistical tests,
focusing on both item analysis and on the entire test. The results indicate
that FCI is a reliable and discriminating tool in most cases. As there are
exceptions, statistical tests should always be done when FCI is administered
in a new context.
\end{abstract}


\section{Introduction}

Standardised multiple-choice tests can be used as a tool in physics
education to assess student learning. A number of such tests have been
developed covering a range of different domains in physics. One of the most
commonly used test, the Force Concept Inventory (FCI), was introduced by
Hestenes, Wells and Swackhamer \cite{Hestenes1992}. FCI has since then been
used as a tool in assessing the efficiency of a number of developed teaching
methods (see for example \cite{Coletta2007}). FCI is limited to the
understanding of the concept of force, but an increased understanding of
force should work as a more general indication of learning in mechanics as a
whole. Considering the extended use of FCI, it should also work as a tool in
assessing the learning of Norwegian students in their introduction physics
courses. In order to investigate the reliability and discrimination power of
FCI at a Norwegian university, a number of statistical tests focusing both
on individual items and on the test as a whole, has been performed. There
exist two aspects of test reliability; consistency and discriminatory power.
A test is said to be reliable if it is consistent within itself and over
time. If a test is shown to be reliable, one can assume that the same
students would get the same score if they would take the test again after a
period of time. A large variance in the test score of a reliable test will
then depend on a systematic variation in the student population, where
different levels of understanding or mastery will give different scores on
the test. Both these aspects of test reliability can be assessed
statistically. In order to evaluate the reliability of the FCI in a
Norwegian context, the test was administrated to different student groups at
the Norwegian University of Science and Technology (NTNU) in Trondheim. Even
if the test is intended to be of general use, the level of the students
understanding or mastery will affect the usefulness of the test, especially
when the group has a higher degree of understanding or mastery.

\section{Background}

A concept inventory is a criterion-referenced test designed to evaluate if
students have an accurate knowledge of a specific set of concepts within a
defined area. Concept inventories are typically organized as multiple-choice
tests in order to ensure that they are objectively scored in a reproducible
manner and possible to administrate in large classes. Unlike a teacher-made
multiple-choice test, questions and response choices in concept inventories
are a subject of extensive research and development. The aims of the
research may include ascertaining (a) the range of what individuals think a
particular question is asking and (b) the most common responses to the
questions. In the concept inventory, each question includes one correct
answer and several distractors. The distractors are incorrect answers that
are usually (but not always) based on students' commonly held
misconceptions. Ideally, the scores should reflect the amount of content
knowledge students has mastered. The purpose of a criterion-referenced test
is to ascertain whether students master a predetermined amount of content
knowledge. The distractors are often based on ideas commonly held by
students, as determined by years of research on misconceptions.

The Force Concept Inventory (FCI) \cite{Hestenes1992} is a multiple-choice
test, designed to assess student understanding of the most basic concepts in
Newtonian physics, particular forces. The test has 30 questions covering six
areas of understanding: kinematics, Newton's First, Second, and Third Laws,
the superposition principle, and types of forces (such as gravitation,
friction). Each question has only one correct Newtonian answer, with
distractors based on student's common misconceptions. A low score indicates
that the student has an Aristotelian view while a high score (typically
around 60\% correct or higher) indicates a Newtonian understanding. The
Norwegian version of FCI used,  was translated and developed by Angell and
collaborators at University of Oslo \cite{Angell2012}.

\section{Student groups}

The test was given in three different courses with different student groups,
both as a pre-(instruction) and post-(instruction) test in the Fall semester
2012. The Courses were traditional calculus-based introductory physics
courses. As all engineering students at NTNU have to take at least one
course in physics, it was possible to administer FCI to both physics masters
and non-physics masters. However, different physics courses are given to
different masters programs, but all courses contain about the same amount of
content relevant for the FCI survey during lectures and are using the same
textbook as the main source. The three groups consisted of students in
different physics courses; Mechanical Physics (TFY4145/FY1001) for Physics
Masters; Physics (TFY4104) for Master students in Marine Technology,
Industrial Economics and Technology Management and Mechanical Engineering;
and Physics (TFY4115) for Master students in Electronics, Engineering
Cybernetics and Nanotechnology. It should also be noted that TFY4104 and
TFY4115 include electromagnetics and thermodynamics, respectively, in
addition to mechanics.

The test was voluntary with no extra credit given. The numbers of students
taking the tests are given in Table \ref{TableKey1}. The result of the tests
with respect to understanding will be presented elsewhere as we focus on the
reliability of the test in this paper. The students in the different groups
have a similar background, but one can assume that the Physics Masters has a
more explicit interest and knowledge in physics and will subsequently score
higher on the FCI. The Physics masters and Nanotechnology students are
generally believed to be high-achieving students as admission grades are
higher compared with the other Master programs. The Physics masters and
Nanotechnology students are first year students while the others are second
year students. By examining the results in the different groups it is
possible to establish the reliability within each group. Using the data from
the individual groups we performed five statistical tests: three focusing on
individual items (item difficulty index, item discrimination index , item
point biserial coefficient) and two on the test a whole (Kuder-Richardson
test reliability and test Ferguson's $\delta$).

\begin{table}[tbp] \centering%
\begin{tabular}{|c|c|c|}
\hline
& Pre-test & Post-test \\ \hline
TFY4104 & 182 & 105 \\ \hline
TFY4115 & 91 & 58 \\ \hline
TFY4145 & 140 & 91 \\ \hline
\end{tabular}
\caption{Number of students taking the FCI.}\label{TableKey1}%
\end{table}%

\section{Item difficulty index}

The item difficulty index (P) is a measure of the difficulty of each test
item and of the test as a whole. It is defined as the ratio of the total
number $N_{1}$ of correct answers to the total number $N$ of students who
answered the specific item:

\begin{equation}
\mathbf{P=}\frac{N_{1}}{N}   \label{1}
\end{equation}

The difficulty index is, however, somewhat misnamed, since it is simply the
proportion of correct answers to a particular item, where the name
\textquotedblleft easiness index\textquotedblright\ would be more
appropiate. The greater P value, the higher percentage of correct answers
and consequently the easier the item is for the population. The difficulty
index will thus depend on the population, something which is the case in
this study. There are a number of different criteria for acceptable values
of the difficulty index for a test \cite{Doran1980}. The optimum value for
an item should be $P=0.5$, while it is useful to have a sensible range. A
widely adopted criterion requires the difficulty index to be between 0.3 and
0.9 for each question. For a test with a large number ($M$) of items it is
more sensible to consider the test difficulty as the average difficulty
index ($\bar{P}$) of all the items ($P_{i}$):

\begin{equation}
\mathbf{\bar{P}=}\frac{1}{M}\sum P_{i}
\end{equation}

\begin{figure}[htbp]
\centering

  \includegraphics[bb=0 0 343 193,width=14.4cm,height=8.1cm,keepaspectratio]{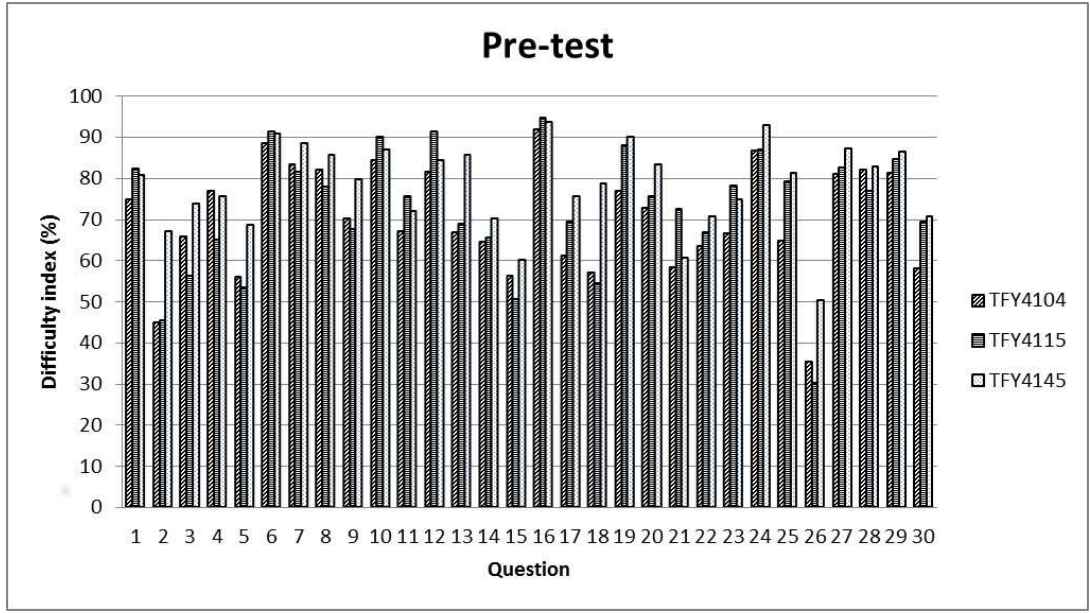}

\caption{Difficulty index pre-test}
\label{fig:Figure1}
\end{figure}

\begin{figure}[htbp]
\centering

  \includegraphics[bb=0 0 343 204,width=11cm,height=6.55cm,keepaspectratio]{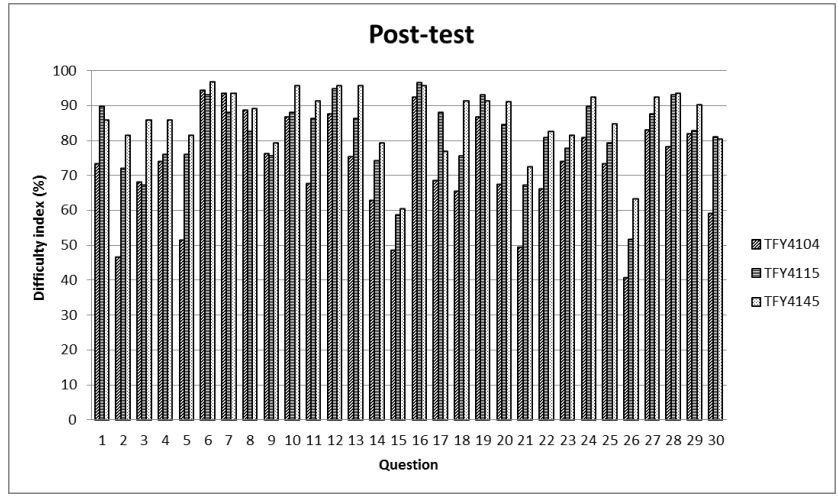}

\caption{Difficulty index post-test}
\label{fig:Figure2}
\end{figure}

Figures \ref{fig:Figure1} and \ref{fig:Figure2} plots the difficulty index $P
$ values for each question in FCI, for the three different student groups
and the pre-test and post-test respectively. The difficulty index in the
pre-tests, fall, in most cases, within the desired range of 0.3-0.9. There
are maximum 4 items with difficulty index above 0.9 in the pre-test,
something that is acceptable. In the post-test the number of questions with
a difficulty above 0.9 rises to 14 and 7, for the TFY4145 and TFY4115
groups, respectively. The average difficulty indexes for the tests are given
in table \ref{TableKey2}. The average difficulty indexes range from 0.70 to
0.78 in the pre-test to 0.72 to 0.86 in the post-test. Even if the results
fall within the acceptable range, the values are very close to the limit of
the acceptable range. Taking in to account that the number of items with a
difficulty index over 0.9 is large, the use of FCI in its original form as a
post-test for a student group such as the TFY4145 group, is very
questionable. However, if one want to study the weaker part of the student
population, the test can still be used.

\bigskip

\bigskip

\begin{table}[tbp] \centering%
\begin{tabular}{|c|c|c|}
\hline
& Pre-test (\%) & Post-test (\%) \\ \hline
TFY4104 & 70 & 72 \\ \hline
TFY4115 & 72 & 81 \\ \hline
TFY4145 & 78 & 86 \\ \hline
\end{tabular}
\caption{Average difficulty index}\label{TableKey2}%
\end{table}%

\section{Item discrimination index}

The item discrimination index (D) is a measure of the discriminatory power
for the individual items in a test. That is, the extent to which an
individual test item distinguishes a student who know the material well from
those who do not. A high discrimination index will indicate a higher
probability for students with a higher level of knowledge to answer the item
correctly, while those with less knowledge will get the wrong answer. The
item discrimination index (D) is calculated by dividing the sample into two
groups of equal size, a high (H) score group and a low (L) score group based
on their individual total scores on the test. For each specific item, one
counts the number of correct answers in both the high and low groups: $N_{H}$
and $N_{L}$. Using  the total number of students taking the test (N), the
discrimination index for a specific item can be calculated as

\begin{equation*}
D=\frac{N_{H}-N_{L}}{N/K}
\end{equation*}

where K is a numerical factor based on how the division into the high and
low group is made. If we split the sample in two, using the median, the high
and low groups consist each of 50\% of the total sample, giving K=2.
However, it is possible to use other groupings, for example taking the top
25\% as the high group and the bottom 25\% as the low group. The 50\%-50\%
grouping may underestimate the discrimination power, since it takes all
students into account even those where the difference is small. To reduce
the probability of underestimating the discrimination power we use a
25\%-25\% grouping. The discrimination index is then expressed as:

\begin{equation*}
D=\frac{N_{H}-N_{L}}{N/4}
\end{equation*}

The range of the item discrimination index D is [-1,+1], where +1 is the
best value and -1 the worst. In the case where all students in the high
score group and none in the low score group get the correct answer the
discrimination index would be +1. If none in the high score group and all in
the low score group get the correct answer the discrimination index would be
-1. These extremes are very unlikely, but shows that items  with a negative
discrimination index should be removed. A question is typical considered to
provide a good discrimination if D$>$0.3 \cite{Doran1980}, lower values
indicate that students resort to guessing on that item. In a test with a
large number of items it is possible to allow a few items with a lower
discrimination index, but the majority should have higher discrimination
indices in order to ensure that the test can distinguish students with
strong and weak mastery. It is useful to calculate the averaged
discrimination index ($\bar{D}$) for all items in the test.

\begin{equation*}
\bar{D}=\frac{1}{M}\sum D_{i}
\end{equation*}

\begin{figure}[htbp]
\centering

  \includegraphics[bb=0 0 343 193,width=14.4cm,height=8.1cm,keepaspectratio]{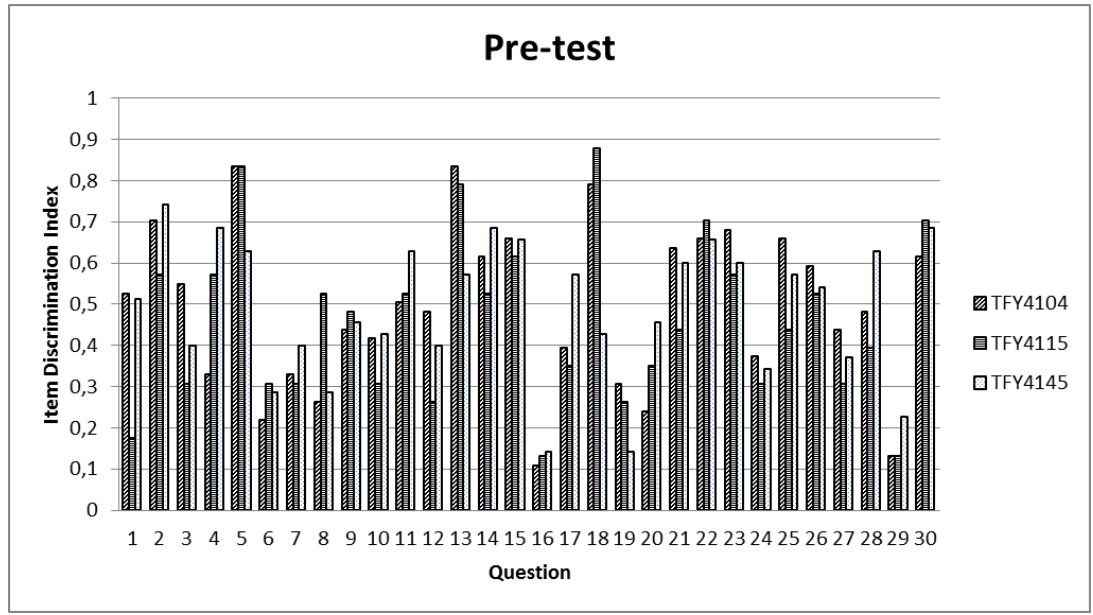}

\caption{Item Discriminating index pre-test}
\label{fig:Figure3}
\end{figure}

\begin{figure}[htbp]
\centering

  \includegraphics[bb=0 0 326 193,width=14.4cm,height=8.53cm,keepaspectratio]{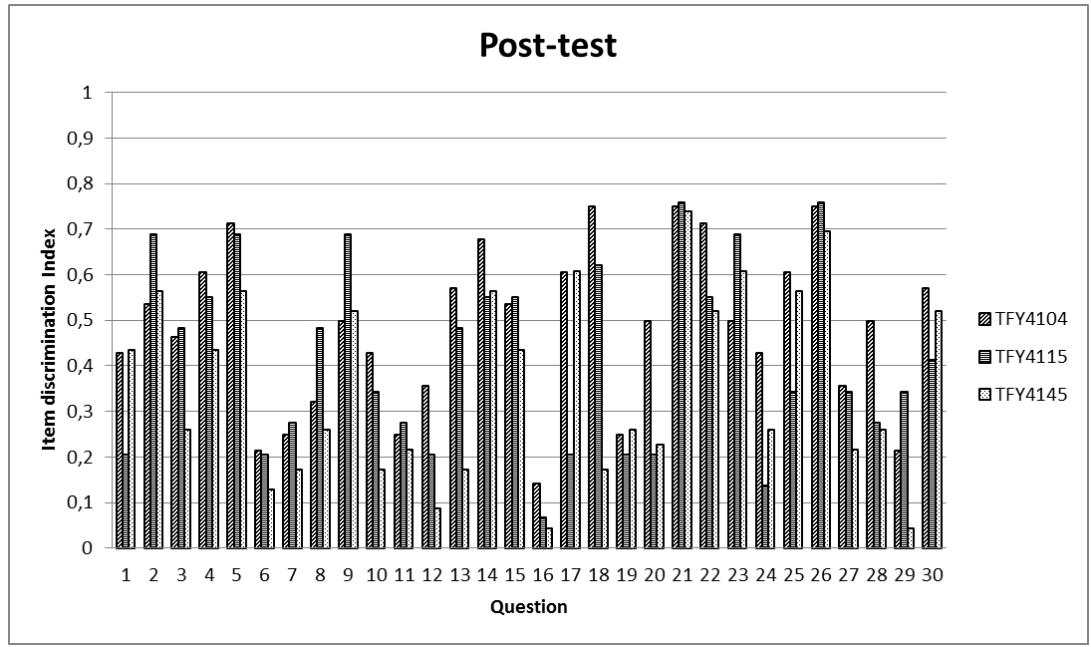}

\caption{Item Discrimination index post-test}
\label{fig:Figure4}
\end{figure}

Figures \ref{fig:Figure3} and \ref{fig:Figure4} plots the discrimination
index D values for the items in FCI, for the three different student groups
and the pre-test and post-test respectively. A majority of the items in the
pre-test has a discrimination index D$>$0.3, only a few has a lower value,
with variations between different groups. The average discrimination indices
are 0.49, 0.49 and 0.45, for the different student groups (TFY4145, TFY4104,
and TFY4115, respectively). This indicates that the FCI has a good
discriminating power in the pre-test situation. In the post-test the number
of questions with the discrimination index D$<$0.3, rises to 15 and 11, out
of 30, for the TFY4145 and TFY4115 groups, respectively, while the averaged
discrimination index decreases to 0.36 and 0.42, respectively. This raises
serious doubts as how applicable the post-test is for the TFY4145 group. The
discrimination power for the TFY4115 group is lower than in the pre-test but
still within the accepted range. In the TFY4104 group, the discrimination
index remains almost the same (0.48). Questions 6, 16 and 29, and to some
extent question 19 are especially doubtful as they combine a high difficulty
index, that is being quite simple, with a low discrimination index, not
distinguishing the high score and low score groups in these student groups.

\section{Point biserial coefficient}

The point biserial coefficient is another measure of the individual item
reliability. It reflects the correlation between the total score and the
score on individual items in the test. A positive coefficient indicates that
a student with a high total score is more likely to answer the item
correctly than a student with a low total score. Thus giving a complementary
measure to the item discrimination index. In order to calculate the point
biserial coefficient for an item, one obtain the correlation between the
score for a question and the total scores. If the number of items in the
test is sufficiently large, $>$20, the test can be viewed as continuous. The
point biserial coefficient can then be defined as:

\begin{equation*}
r_{pbc}=\frac{\bar{X}_{1}-\bar{X}_{0}}{\sigma_{x}}\sqrt{\frac{P}{1-P}}
\end{equation*}

Where $\bar{X}_{1}$ is the average total score for those who answered a item
correctly, $\bar{X}_{0}$ is the average total score for all participants, $%
\sigma _{x}$ is the standard deviation of the total scores and P is the
difficulty index for this specific item. For an item to be considered as
reliable it should be consistent with the whole test,  a high correlation
between individual item scores and the total score is desirable. A
satisfactory point biserial coefficient is $r_{pbc}$$>$ 0.2\cite{Doran1980}.
Items with lower values may be used, as long as the number of these items is
small, but the test as a whole should have an average higher than 0.2. The
average point biserial coefficients for the different student groups are
given in table \ref{TableKey3}. All values are greater than 0.2 so the
overall items has a fairly high correlation with the whole test. Figures \ref%
{fig:Figure5} and \ref{fig:Figure6} shows the point biserial coefficients
for the individual items in the pre- and post-tests for different student
groups, respectively. It should be noted that questions 16, 19 and 29
overall show a lower degree of correlation than the others. As these
questions also show a lower degree of discrimination and these might be
subject to revision, at least in the context of the student groups in this
study. There is a course-dependent variation of the point biserial
coefficient for the post-test. These variations might be due to statistical
variations or different course context.

\begin{figure}[htbp]
\centering

  \includegraphics[bb=0 0 343 193,width=14.4cm,height=8.1cm,keepaspectratio]{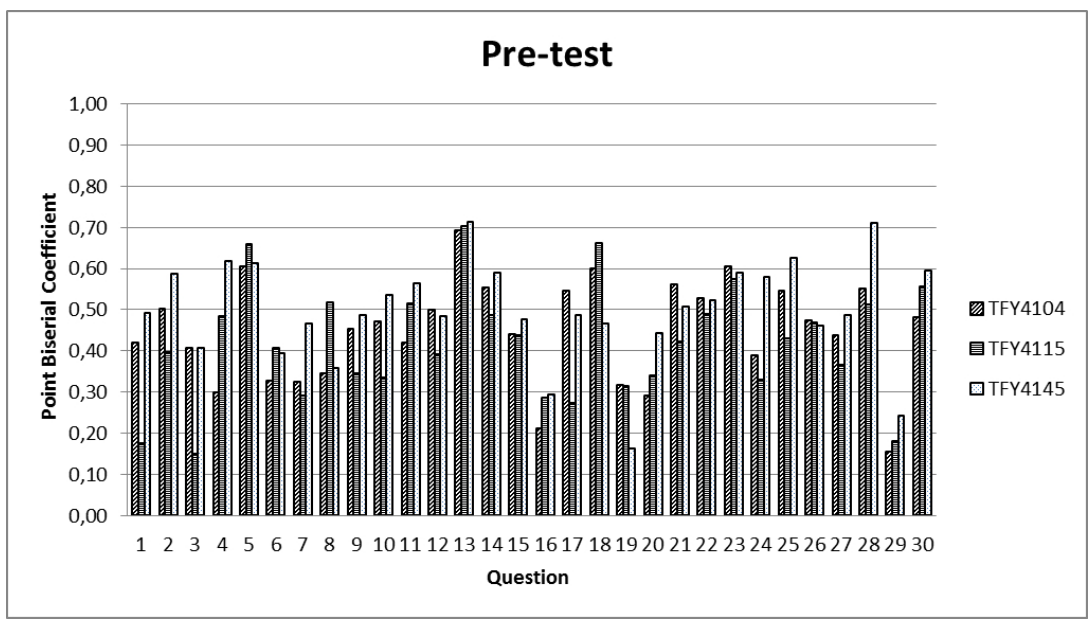}

\caption{Point biserial coefficient pre-test}
\label{fig:Figure5}
\end{figure}

\begin{figure}[htbp]
\centering

  \includegraphics[bb=0 0 343 204,width=14.4cm,height=8.56cm,keepaspectratio]{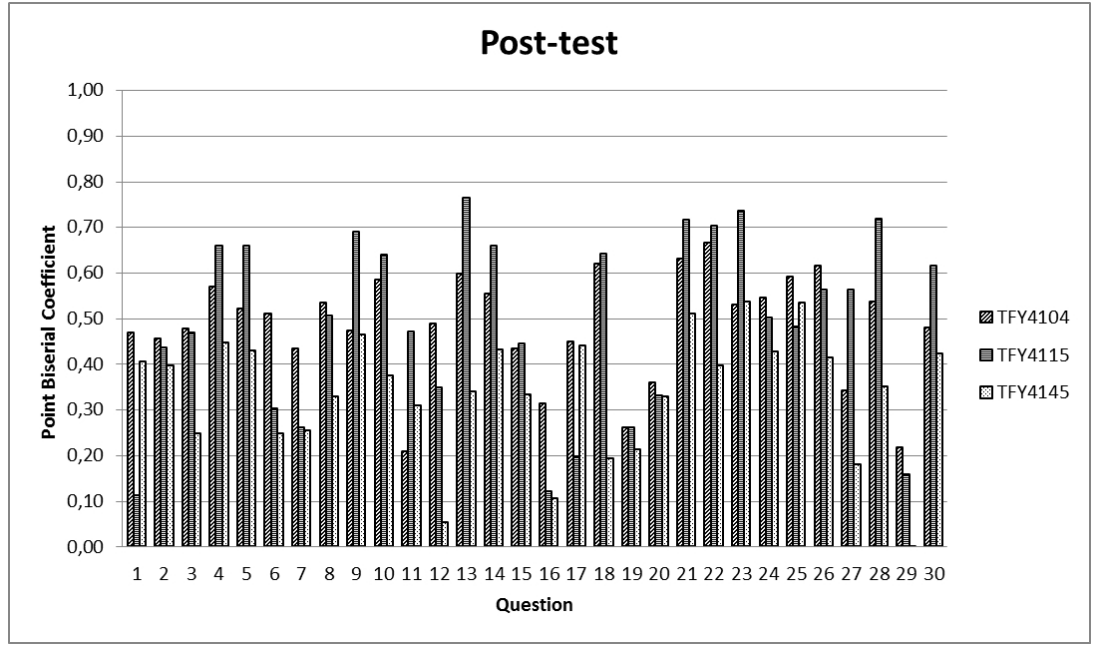}

\caption{Point biserial index post-test}
\label{fig:Figure6}
\end{figure}

\begin{table}[tbp] \centering%
\begin{tabular}{|c|c|c|}
\hline
& Pre-test & Post-test \\ \hline
TFY4104 & 0.45 & 0.48 \\ \hline
TFY4115 & 0.42 & 0.49 \\ \hline
TFY4145 & 0.50 & 0.34 \\ \hline
\end{tabular}
\caption{Average point biserial coefficients}\label{TableKey3}%
\end{table}%

\section{Test analysis}

The reliability of single items in the test is measured by the point
biserial coefficient. In order to examine the reliability of the test as a
whole, other methods have to be used. In this work we use two measures of
the reliability for the test as a whole: Kuder-Richardson reliability index
and Ferguson's delta ($\delta$).

\subsection{Kuder-Richardson reliability index}

A not very practical way to evaluate the reliability of a test, is to
administer it twice to the same sample. In such a case we would expect a
significant correlation between the two test scores, provided the students'
performance is stable and the test conditions are the same. The correlation
coefficient between the two sets of scores will be defining the reliability
index of the test. It is obvious that this method is not practical to use.
In the case of a test that has been designed specifically for a certain
knowledge domain and with parallel questions, the Spearman-Brown formula 
\cite{Ghiselli1981} can be used to calculate the reliability index. This
equation connects the reliability index with the correlation between any
parallel equally sized subsets in the test. Kuder and Richardson \cite%
{Kuder1937} developed this idea further by dividing the test into the
smallest possible subsets, that is individual items. This means that each
item is considered as a single parallel test, and assuming that the means,
variance and standard deviation is the same for all items in the whole test.
The result derived gives the reliability index as:

\begin{equation*}
r_{test}=\frac{M}{M-1}\left( 1-\frac{\sum\sigma_{xi}^{2}}{\sigma_{x}^{2}}%
\right) 
\end{equation*}

where M is the number of items in the whole test, $\sigma _{xi}$ is the
standard deviation for the ith item score and $\sigma _{x}$ is the standard
deviation of the total test score. This expression takes the different
variances of the individual items into account, relaxing the assumption that
all items must have the same means, variance and standard deviation. For
multiple-choise tests the formula can be rewritten as:

\begin{equation*}
r_{test}=\frac{M}{M-1}\left( 1-\frac{\sum P_{i}(1-P_{i})}{\sigma_{x}^{2}}%
\right) 
\end{equation*}

where M is the number of items in the test, $P_{i}$ is the difficulty index
for each item and $\sigma _{x}$ is the standard deviation of the total test
score. These are the Kuder-Richardson reliability formulas, often referred
to as KR-20 and KR-21 as being formula 20 and 21 in Kuder and Richardson's
original paper \cite{Kuder1937} The possible range of the Kuder-Richardson
reliability index is between 0 and 1, where a value greater than 0.7 would
make the test reliable for group measurements and a value over 0.8 for
assessing individuals \cite{Doran1980}. In this study the obtained
Kuder-Richardson reliability indices are all over 0.8 (Table \ref{TableKey4}%
).Something that also open up for individual assessment.

\bigskip%
\begin{table}[tbp] \centering%
\begin{tabular}{|c|c|c|}
\hline
& Pre-test & Post-test \\ \hline
TFY4104 & 0.87 & 0.88 \\ \hline
TFY4115 & 0.84 & 0.86 \\ \hline
TFY4145 & 0.90 & 0.87 \\ \hline
\end{tabular}
\caption{Kuder-Richardson reliability index}\label{TableKey4}%
\end{table}%

\subsection{Ferguson's delta}

Ferguson's delta is another widely used whole test statistic. It measures
the discriminatory power of the whole test by investigating how the
students' individual scores are distributed. In a test one aims at a broad
distribution of total scores, as this is supposwed to show a better
discrimination. The expression of Ferguson's delta can be written as \cite[p
150]{Kline1986}:

\begin{equation*}
\delta=\frac{N^{2}-\sum f_{i}^{2}}{N^{2}-\left( N^{2}/\left( M+1\right)
\right) }
\end{equation*}

where N is the number of students taking the test, M is the number of items
in the test and $f_{i}$ is the frequency of cases with the same score. One
should be aware that Ferguson's delta is more a measure of the population
than the test itself, since a change in population will change the result of
the Ferguson's delta formula, while not testing the test itself.  If a test
and population combined has a Ferguson's delta greater than 0.90, it is
considered to provide a good discrimination for this population \cite[p 144]%
{Kline1986}. In our study the Ferguson's delta is greater than 0.90, in all
cases as is shown in table \ref{TableKey5}.

\bigskip%
\begin{table}[tbp] \centering%
\begin{tabular}{|c|c|c|}
\hline
& Pre-test & Post-test \\ \hline
TFY4104 & 0.98 & 0.97 \\ \hline
TFY4115 & 0.97 & 0.94 \\ \hline
TFY4145 & 0.96 & 0.91 \\ \hline
\end{tabular}
\caption{Ferguson's delta}\label{TableKey5}%
\end{table}%

\section{Discussion}

The reliability and discriminatory power of the Force Concept Inventory has
been evaluated using five statistical tests in three different student
groups, both in pre-instructional and post-instructional tests, at the
Norwegian University of Science and Technology (NTNU) . The aim of this
study was to test the applicability of the FCI in different contexts, as
made possible with Physics majors and engineering students required to take
at least one physics course at NTNU. We have found that the FCI is reliable
and discriminating enough for pre-tests in all student groups. The post-test
for Physics majors (TFY4145) can not be considered as applicable in the
present form for the full group, the average difficulty index (86\%) has
reached a level where ceiling effects will cause problems. The average
discrimination index, though still over the 0.3 level, is not a good
indicator as half of the questions have a discrimination index below that
level. A similar but not as serious problem can also be seen in the TFY4115
group. However, it is still possible to use the test for specific subgroups,
that is low achieving students, in order to investigate their understanding.
The Force Concept Inventory is a widely used instrument, but as has been
shown here, it can not be used without taking the context and student groups
into account. Used on a high-achieving group, there is a substantial risk of
encountering ceiling effects, with a decrease in discriminatory power. It
will still be useful for the students within this group that has not
obtained an understanding of the fundamental concepts. Questions 6, 16, 19
and 29 in the FCI are somewhat problematic and might be replaced with other
questions in a high-achieving group, such as TFY4145. However, one might
also consider constructing a special high-achieving FCI suitable for Physics
majors.


\end{document}